\documentclass[a4paper]{article}
\usepackage[latin1]{inputenc}
\usepackage[T1]{fontenc}
\usepackage[english]{babel}
\usepackage{amsthm,amsfonts,amsmath}
\usepackage{proof}
\usepackage{color}
\usepackage{geometry}
\usepackage{graphicx}
\usepackage{bbold}
\usepackage{pgfplots}
\usepackage{multirow}
\usepackage{scalefnt}
\usepackage{eurosym}
\usepackage{colortbl}
\usepackage{caption}
\usepackage{subcaption}
\usepackage{authblk}
\usepackage{natbib}

\definecolor{llightgray}{rgb}{0.85,0.85,0.85}

\def\R{\mathbb{R}}
\def\N{\mathbb{N}}
\def\V{\mathbb{V}}
\def\E{\mathbb{E}}

\def\Ind{\mathbb{1}}
\DeclareMathOperator{\cov}{Cov}

\DeclareMathOperator{\Tr}{Tr}
\DeclareMathOperator{\arginf}{arginf}

\newtheorem{theorem}{Theorem}
\newtheorem{remark}{Remark}

\title{New improvements in the use of dependence measures for sensitivity analysis and screening}
\author[1]{Matthias De Lozzo}
\author[1]{Amandine Marrel}
\affil[1]{CEA,DEN,DER, F-13108 Saint Paul Lez Durance, France}

\date{}

\begin{document}
\maketitle

\begin{abstract}

Physical phenomena are commonly modeled by numerical simulators. Such codes can take as input a high number of uncertain parameters and it is important to identify their influences via a global sensitivity analysis (GSA). However, these codes can be time consuming which prevents a GSA based on the classical Sobol' indices, requiring too many simulations. This is especially true as the number of inputs is important. To address this limitation, we consider recent advances in dependence measures, focusing on the distance correlation and the Hilbert-Schmidt independence criterion (HSIC). Our objective is to study these indices and use them for a screening purpose.

Numerical tests reveal some differences between dependence measures and classical Sobol' indices, and preliminary answers to ``What sensitivity indices to what situation?'' are derived. Then, two approaches are proposed to use the dependence measures for a screening purpose. The first one directly uses these indices with independence tests; asymptotic tests and their spectral extensions exist and are detailed. For a higher accuracy in presence of small samples, we propose a non-asymptotic version based on bootstrap sampling. The second approach is based on a linear model associating two simulations, which explains their output difference as a weighed sum of their input differences. From this, a bootstrap method is proposed for the selection of the influential inputs. We also propose a heuristic approach for the calibration of the HSIC Lasso method. Numerical experiments are performed and show the potential of these approaches for screening when many inputs are not influential.
\end{abstract}

\vspace{0.5cm}


\section{Introduction}

Numerical simulators are widely used in the industry for the representation of physical phenomena \citep{santner}. Such models take as input a high number of numerical and physical explanatory variables. The information on these underlying input parameters is often limited or uncertain. Commonly, the uncertainties on the input parameters are modeled by probabilistic distributions. Then, the objective is to assess how these uncertainties can affect the model output. For this, computer experiments methodologies based upon statistical advanced techniques are useful \citep{Rocquigny,kleijnen2007design}.

Sensitivity Analysis (SA) methods allow to answer the question ``How do the input parameters variations contribute, qualitatively or quantitatively, to the variation of the output?'' \citep{saltelli}. More precisely, these tools can detect non-significant input parameters in a screening context, determinate the most significant ones, measure their respective contributions to the output or identify an interaction between several inputs which impacts strongly the model output. In such a way, engineers can guide the characterization of the model by reducing the output uncertainty: they can calibrate the most influential inputs and fix the non-influential ones to nominal values. Many surveys on SA exist in the literature, such as \citet{kleijnen1997}, \citet{frey} or \citet{Helton20061175}; they divide the SA into two sub-domains: the Local Sensitivity Analysis (LSA) and the Global Sensitivity Analysis (GSA). The first one studies the effects of small input perturbations around nominal values on the model output. Usually this deterministic approach considers the partial derivatives of the model at a specific value of the input vector \citep{cacuci}. The second sub-domain of SA considers the impact of the input uncertainty on the output over the whole variation domain of uncertain inputs, that is why it is called Global SA \citep{saltelli:2008}. 

The GSA can be used for quantitative or qualitative purposes, specific tools being dedicated to each aim. From one hand, quantitative GSA methods supply an order of the input parameters which is function of their dependence to the output.  Among them, the Derivative-based Global Sensitivity Measures (DGSM) consider the mean of the model gradient over the whole input domain \citep{lamboniDGSM}, not at a specific point like in LSA \citep{cacuci}. Another approach is based on the decomposition of the output variance; in particular, the Sobol' indices are widely used and measure the proportion of the output variance explained by each input parameter \citep{sobol}.  Other authors propose to consider all the probabilistic distribution and not only the variance, comparing the distribution of the output conditioned by an input parameter with the unconditioned one \citep{Borgonovo2007771}. 

From the other hand, qualitative GSA uses less cost\-ly tools coming from the screening field. These methods can detect the input-output dependences and separate the input parameters into two groups: the non-sig\-ni\-fi\-cant and the significant ones. Despite of the criticisms with respect to the underlying hypotheses \citep{saltelliOAT}, the basic screening tool is the one-at-a-time (OAT) design which consists in changing the values of each input parameter in turn from a control level scenario to a lower or upper level and measuring the evolution magnitude of the output \citep{OAT}. Another method is the Morris design which consists in the repetition of many OAT designs, in order to get a mean value and a standard deviation for each input elementary effect \citep{morris}. Other screening methods are currently used, such as the sequential bifurcation in a sparse context, when the number of significant input parameters is considerably lower than the total one which is greater than the number of observations \citep{seqbif}. When the number of observations and the number of input parameters are of the same order, factorial fractional designs and other popular designs of experiments can be applied \citep{montgomery}. Very recently, the use of Sobol' indices for sparse problems has been investigated \citep{decastrojanon}, in a screening framework where the effective dimension is much lower than the number of input parameters \citep{effectivedimension}.\\

Among all these GSA methods, the quantitative ones like Sobol' indices give a more accurate information about the dependence between the input parameters and the model output, while the qualitative methods are more imprecise. Moreover, Sobol' indices have been applied to many industrial problems in order to reduce the output variance. Nevertheless, these methods require many thousands of computer experiments in order to build reliable estimators of the sensitivity indices. Moreover, the number of required simulations is proportional to the number of inputs so as to preserve the precision of the sensitivity index estimator. Consequently, in the presence of a costly numerical simulator, quantitative GSA can not be performed directly, for high-dimensional problems.

A first alternative consists in replacing the computer code by a surrogate model and computing a quantitative GSA on this model. For example, \citet{marrelgpmsobol} and \citet{Sudret2008964} estimate the Sobol' indices thanks to Gaussian process models and polynomial cha\-os expansions respectively. However, the estimator accuracy depends on the precision of the surrogate model which can be weak if the learning sample is not enough representative. Moreover, the construction of the surrogate model in a high dimensional context (several decades of input parameters) is still an open problem.

Another alternative consists in using cheaper sensitivity indices which are potentially less accurate than the Sobol' ones but easier to compute (smaller CPU time). Qualitative GSA methods previously cited are commonly used to this aim. Nevertheless they often require either strong hypotheses on the model such as linearity, monotony or absence of interactions, or a number of observations much greater than the number of input parameters. The non-respect of these assumptions can lead to incorrect quantitative conclusions. Moreover, many of these screening methods consider specific design of experiments which can not be reused for other studies. Recently, new dependence measures removing these limitations have been developed by statisticians \citep{gretton,dcortest} and applied in genomics, imagery or cross-lan\-gua\-ge information retrieval \citep{HSICapp}. They have been studied in the field of global sensitivity analysis: they seem more robust than Sobol' indices, promising in a screening aim and can provide an information complementary to the Sobol' indices \citep{daveiga}. They can also make easier the metamodel construction by reducing the input number or guiding it in a sequential way; then, a quantitative GSA is performed to obtain an information more accurate on the input parameters identified as significant by the qualitative GSA.\\

\textbf{In this paper, we focus our attention on the use of these new dependence measures for qualitative GSA. We develop several independence tests to use these measures for a screening purpose: ones based on the estimator of the sensitivity index directly, others based on a linear decomposition and model selection methods. We also performed different numerical experiments to study the behavior of the dependence measures and compare the different proposed tests.}\\

Firstly, we present in Section \ref{sec:measures} some dependence measures for the sensitivity of an output with respect to an input parameter. Secondly in Section \ref{sec:stattest}, we deal with asymptotic and non-asymptotic statistical tests based on these dependence measures for feature selection. In Section \ref{sec:bootlinreg}, we propose a linear model associated to these dependence measures and build bootstrap tests and penalized regression techniques. Finally in Section \ref{sec:results}, numerical experiments are tested on analytical models, starting with a questioning around the meaning and the complementarity of the different sensitivity indices: ``What sensitivity indices to what situation?''.

\section{Dependence measures for an input-output relation}\label{sec:measures}

We consider a computer code $Y=f\left(X_1,\ldots,X_d\right)$ who\-se output $Y$ and input parameters $X_1,\ldots,X_d$ belong to some measurable spaces $\mathcal{Y}$, $\mathcal{X}_1$, ..., $\mathcal{X}_d$. We note $X=\left(X_1,\ldots,X_d\right)$ the input vector of $f$. $\mathcal{Y}$ and $\mathcal{X}_k$ are commonly equal to $\R$, for all $k\in\{1,\ldots,d\}$, but sometimes engineers are in front of more complex situations where $X_k$ or $Y$ can be a vector, a time- or a space-discretized function, and so on. The $d$ input parameters are considered as random variables whose laws are perfectly known. Consequently, the output $Y$ is also a random variable whose probability distribution is usually unknown and unapproachable because of the curse of dimensionality. We present in the following some measures of the dependence between an input parameter $X_k$ and the output $Y$ of the model $f$. The associated estimators are built using $(X_{1,i},\ldots,X_{d,i},Y_i)_{1\leq i \leq n}$, a $n$-sample of $(X_1,\ldots,X_d,Y)$.

	\subsection{The Pearson's and Spearman's correlation coefficients}\label{sec:pearspea}

First of all, we can cite naive importance measures such as Pearson's and Spearman's correlation coefficients \citep[see, e.g.,][]{kendall}. These quantities evolve in the interval $[-1,1]$, reaching the bounds for a total correlation between the variables $X_k\in\mathcal{X}_k\subset\R$ and $Y\in\mathcal{Y}\subset\R$ and equaling zero for an absolute uncorrelation. The Pearson's one has the well-known formulation	
$$\rho(X_k,Y)=\frac{\cov(X_k,Y)}{\sqrt{\V[X_k]\V[Y]}}$$
and is estimated by 
$$\rho_n(X_k,Y)=\frac{ \sum_{i=1}^n (X_{k,i}-\bar{X}_k)(Y_i-\bar{Y}) }{\sqrt{ \sum_{i=1}^n (X_{k,i}-\bar{X}_k)^2\sum_{j=1}^n (Y_j-\bar{Y})^2}}$$
where $\bar{X}_k=n^{-1}\sum_{i=1}^nX_{k,i}$ and $\bar{Y}=n^{-1}\sum_{i=1}^nY_i$.
The Spearman's correlation coefficient is a version of the Pearson's one applied on the ranks of $(X_{k,i},Y_i)_{1\leq i\leq n}$:
$$\rho^{(S)}_n(X_k,Y)=1-\frac{6\sum_{i=1}^nd_i^2}{n(n^2-1)}$$
where $d_i=r(X_{k,i})-r(Y_i)$ is the difference between the ranks of $X_{k,i}$ and $Y_i$. Asymptotically, the associated statistics $t_n=\rho_n(X_k,Y)\sqrt{\frac{n-2}{1-\rho_n^2(X_k,Y)}}$ and $t^{(S)}_n=\rho_n^{(S)}(X_k,Y)\sqrt{\frac{n-2}{1-\left(\rho_n^{(S)}(X_k,Y)\right)^2}}$ follow a Student distribution with $n-2$ degrees of liberty and significance tests can easily be proposed for feature selection.\\

Despite of their simple formulations, the Pearson's and Spearman's coefficients take into account only linear and monotonous effects respectively. Consequently, they cannot deal with non-monotonic behavior and interactions between input parameters.

	\subsection{The distance correlation}\label{sec:dcov}
	
To address the limitations of correlation coefficients listed in Section \ref{sec:pearspea}, a first dependence measure presented in \citet{daveiga} offers an interesting alternative. This quantity is based on the marginal distributions of the couple $(X_k,Y)$ and avoids making parametric assumptions on the model $Y=f(X)$. Considering the random variables $X_k\in\mathcal{X}_k\subset\R^{d_k}$ and $Y\in\mathcal{Y}\subset\R^p$ with characteristic functions $\Phi_{X_k}$ and $\Phi_Y$, the distance covariance is defined by

\begin{eqnarray}
\mathcal{V}^2(X_k,Y)=\int_{\R^{d_k+p}}&&|\Phi_{X_k,Y}(t,s)-\Phi_{X_k}(t)\nonumber\\
&\times&\Phi_Y(s)|^2w(t,s)dtds
\label{eq:distcov}
\end{eqnarray}	
\noindent where $w(t,s)=(c_{d_k}c_p\|t\|_2^{1+d_k}\|s\|_2^{1+p})^{-1}$ with the constants $c_l=\pi^{(1+l)/2}/\Gamma((1+l)/2)$ for $l\in\N$ and $\|.\|_2$ is the $L^2$ norm \citep{dcortest}. This quantity $\mathcal{V}^2(X_k,Y)$ is equal to zero if and only if the characteristic function $\Phi_{X_k,Y}$ of the couple $(X_k,Y)$ is equal to the product of $\Phi_{X_k}$ and $\Phi_Y$, that is to say only and only if $X_k$ and $Y$ are independent. In other words, the distance covariance is a good indicator of the dependence between $X_k$ and $Y$, without any hypothesis on the law of $X_k$ or the type of relation between $X_k$ and $Y$.\\

This distance covariance (\ref{eq:distcov}) can be expressed in terms of Euclidean distances:
\begin{eqnarray*}
\mathcal{V}^2(X_k,Y)&=&\E_{X_k,X_k',Y,Y'}\left[\|X_k-X_k'\|_2\|Y-Y'\|_2\right]\\
&+&\E_{X_k,X_k'}\left[\|X_k-X_k'\|_2\right]\E_{Y,Y'}\left[\|Y-Y'\|_2\right]\\
&-&2\E_{X_k,Y}\left[\E_{X_k'}\left[\|X_k-X_k'\|_2\right]\E_{Y'}\left[\|Y-Y'\|_2\right]\right]
\end{eqnarray*}
where $(X',Y')$ is an independent and identically distributed copy of $(X,Y)$ and where $\E_Z$ represents the statistical mean in $Z$, for any random variable $Z$. From this statement, \citet{dcortest} propose an estimator of $\mathcal{V}^2(X_k,Y)$:
\begin{eqnarray*}
\mathcal{V}^2_n(X_k,Y)&=&\frac{1}{n^2}\sum_{i,j=1}^n\|X_{k,i}-X_{k,j}\|_2\|Y_i-Y_j\|_2\\
&+&\frac{1}{n^2}\sum_{i,j=1}^n\|X_{k,i}-X_{k,j}\|_2\sum_{i,j=1}^n\|Y_i-Y_j\|_2\\
&-&\frac{2}{n^3}\sum_{i=1}^n\left[\sum_{j=1}^n\|X_{k,i}-X_{k,j}\|_2\sum_{j=1}^n\|Y_i-Y_j\|_2\right]
\label{eq:distcovest}
\end{eqnarray*}

\noindent which can be rewritten in a more compact form:
$$\mathcal{V}^2_n(X_k,Y)=\frac{1}{n^2}\Tr\left[G^{(X_k)}HG^{(Y)}H\right]$$
where $H=\left(\delta_{ij}-\frac{1}{n}\right)_{1\leq i,j \leq n}$ is a centering matrix and where $G^{(X^{(k)})}$ and $G^{(Y)}$ are the Gram matrices defined by $G^{(X_k)}=(\|X_{k,i}-X_{k,j}\|_2)_{1\leq i,j \leq n}$ and $G^{(Y)}=(\|Y_i-Y_j\|_2)_{1\leq i,j \leq n}$.\\

An estimator more efficient from a computational point of view, but with a less straightforward formulation, is also proposed in \citet{dcortest}:
\begin{equation}
\mathcal{V}^2_n(X_k,Y)=\frac{1}{n^2}\sum_{i,j=1}^nA_{ij}B_{ij}
\label{eq:distcovestAB}
\end{equation}
where $A_{ij}=G_{ij}^{(X_k)}-\bar{G}_{i.}^{(X_k)}-\bar{G}_{.j}^{(X_k)}+\bar{G}_{..}^{(X_k)}$ and $B_{ij}=G_{ij}^{(Y)}-\bar{G}_{i.}^{(Y)}-\bar{G}_{.j}^{(Y)}+\bar{G}_{..}^{(Y)}$ with $\bar{M}_{.j}=n^{-1}\sum_{i=1}^nM_{ij}$, $\bar{M}_{i.}=n^{-1}\sum_{j=1}^nM_{ij}$ and $\bar{M}_{..}=n^{-2}\sum_{i=1}^n\sum_{j=1}^nM_{ij}$, for all $M\in\mathcal{M}_n(\R)$.\\

Finally from the distance covariance, the distance correlation $\mathcal{R}^2(X_k,Y)$ is proposed:
\begin{eqnarray}
\mathcal{R}^2(X_k,Y)=\frac{\mathcal{V}^2(X_k,Y)}{\sqrt{\mathcal{V}^2(X_k,X_k)\mathcal{V}^2(Y,Y)}}
\label{eq:distcorr}
\end{eqnarray}
if $\mathcal{V}^2(X_k,X_k)\mathcal{V}^2(Y,Y)>0$ and $0$ otherwise. This normalization involves that $\mathcal{R}^2(X_k,Y)$ is included in the interval $[0,1]$, like the absolute Pearson's correlation coefficient, which makes its interpretation easier. The associated plug-in estimator deduced from (\ref{eq:distcovestAB}) is
$$\mathcal{R}^2_n(X_k,Y)=\frac{\mathcal{V}^2_n(X_k,Y)}{\sqrt{\mathcal{V}^2_n(X_k,X_k)\mathcal{V}^2_n(Y,Y)}}.$$
	
	\subsection{The Hilbert-Schmidt dependence measure}\label{sec:HSIC}
	
Instead of quantifying the link between an input parameter and the model output from an analysis of their characteristic functions, \citet{gretton} propose to use the covariance between some transformations of these random variables. More precisely, we consider the random variables $X\in\mathcal{X}_k$ and $Y\in\mathcal{Y}$, with the probability density functions $p_{X_k}$ and $p_Y$ and where $\mathcal{X}_k$ and $\mathcal{Y}$ are any measurable spaces. We associate to $X_k$ an universal Reproducing Kernel Hilbert-Schmidt space (RKHS) $\mathcal{F}_k$ composed of functions mapping from $\mathcal{X}_k$ to $\R$ and defined by the kernel function $k_{\mathcal{X}_k}$ \citep{RKHS}. The same transformation is realized with $Y$, considering the universal RKHS $\mathcal{G}$ and the kernel function $k_{\mathcal{Y}}$. We note $\langle .,. \rangle_{\mathcal{F}_k}$ and $\langle .,. \rangle_{\mathcal{G}}$ the scalar product over $\mathcal{F}_k$ and $\mathcal{G}$ respectively.

Then, the operator of crossed-covariance $C_{X_kY}$ associated to the probability density function $p_{X_kY}$ of $(X_k,Y)$ is the linear operator mapping from $\mathcal{G}$ to $\mathcal{F}_k$ and defined for all $f\in\mathcal{F}_k$ and for all $g\in\mathcal{G}$ by:
$$\langle f,C_{X_kY}g \rangle_{\mathcal{F}_k}=\text{Cov}\left(f(X_k),g(Y)\right).$$
This operator generalizes the covariance matrix bet\-ween $X_k$ and $Y$. Indeed, thanks to the non-linear kernels which remove hypotheses such as linearity or mo\-no\-to\-ny, it takes into account dependences more complex than the Pearson's and Spearman's coefficients.\\

Finally, the Hilbert-Schmidt Independence Cri\-te\-rion (HSIC) is defined in \citet{gretton} as the Hilbert-Schmidt norm of the operator $C_{X_kY}$ \citep{HSnorm}:
$$\|C_{X_kY}\|_{HS}^2=\sum_{i,j}\langle u_i,C_{X_kY}v_j \rangle_{\mathcal{F}_k}$$
where $(u_i)_{i\geq 0}$ and $(v_j)_{j\geq 0}$ are orthonormal bases of $\mathcal{F}_k$ and $\mathcal{G}$, respectively.

More precisely, we have: 
\begin{eqnarray}
&&\text{HSIC}(X_k,Y)_{\mathcal{F}_k,\mathcal{G}}=\|C_{X_kY}\|_{HS}^2\nonumber\\
&=&\E_{X_k,X_k',Y,Y'}\left[k_{\mathcal{X}_k}(X_k,X_k')k_{\mathcal{Y}}(Y,Y')\right]\nonumber\\
&+&\E_{X_k,X_k'}\left[k_{\mathcal{X}_k}(X_k,X_k')\right]\E_{Y,Y'}\left[k_{\mathcal{Y}}(Y,Y')\right]\nonumber\\
&-&2\E_{X_k,Y}\left[\E_{X_k'}\left[k_{\mathcal{X}_k}(X_k,X_k')\right]\E_{Y'}\left[k_{\mathcal{Y}}(Y,Y')\right]\right].
\label{eq:HSIC}
\end{eqnarray}

Similarly to the distance covariance (\ref{eq:distcorr}), this dependence measure (\ref{eq:HSIC}) is equal to zero if and only if $X_k$ and $Y$ are independent, without emitting any hypothesis about the nature of the relation between $X_k$ and $Y$.\\

From a $n$-sample $(X_i,Y_i)_{1\leq i \leq n}$ of $(X,Y)$, an estimator of the measure $\text{HSIC}(X_k,Y)_{\mathcal{F}_k,\mathcal{G}}$ is proposed in \citet{gretton}:
\begin{eqnarray}
\text{HSIC}_n(X_k,Y)=\frac{1}{n^2}\Tr(K_{\mathcal{X}_k}HK_{\mathcal{Y}}H).
\label{eq:HSICest}
\end{eqnarray}
The Gram matrices $K_{\mathcal{X}_k}$ and $K_{\mathcal{Y}}$ are defined by $K_{\mathcal{X}_k}=\left(k_{\mathcal{X}_k}(X_{k,i},X_{k,j})\right)_{1\leq i,j \leq n}$ and $K_{\mathcal{Y}}=\left(k_{\mathcal{Y}}(Y_i,Y_j)\right)_{1\leq i \leq n}$; $H$ is the centering matrix introduced in the case of the distance covariance. Following the same way as (\ref{eq:distcovestAB}), we propose to reduce the calculation time of (\ref{eq:HSICest}) using $G^{(X_k)}:=K_{\mathcal{X}_k}$ and $G^{(Y)}:=K_{\mathcal{Y}}$. \\

The kernel functions involved in the HSIC definition can belong to various classes of kernel functions, such as the Gaussian, the Laplacian or the Mat\'ern family \citep{srip}. Note that these functions often require hyperparameter values which can be deduced from heuristic processes or fixed in order to maximize the HSIC value \citep{Balasubramanian}. In this paper, we consider the Gaussian kernel function $k(z_i,z_j)=\exp\left(-\sum_{k=1}^{n_z}\frac{\left(z_{k,i}-z_{k,j}\right)^2}{\sigma^2_k}\right)$ for inputs and outputs and $\sigma^2$ is estimated by the empirical variance associated to $z_{k,1},\ldots,z_{k,n}$.

\section{Significance tests for feature selection}\label{sec:stattest}

In a screening context, the objective is to separate the input parameters into two sub-groups, the significant ones and the non-significant ones. For this, we propose to use statistical hypothesis tests based on dependence measures described in Section \ref{sec:measures}. For a given input $X_k$, it aims at testing the null hypothesis ``$\mathcal{H}_0^{(k)}$: $X_k$ and $Y$ are independent'', against its alternative ``$\mathcal{H}_1^{(k)}$: $X_k$ and $Y$ are dependent''. The significance level\footnote{The significance level of a statistical hypothesis test is the rate of the type I error which corresponds to the rejection of the null hypothesis $\mathcal{H}_0$ when it is true.} of these tests is hereinafter noted $\alpha$. Some asymptotic results exist in this domain for the dependence measures; we briefly  present some of them in the following, based on the notations of Section \ref{sec:measures}. In a second part, we develop spectral approximations of the asymptotic laws governing the statistics involved in these tests, which can be useful for medium size samples. Finally, we propose to extend these results to the non-asymptotic case thanks to a bootstrap approach.

	\subsection{Asymptotic tests of independence}\label{sec:asymptest}
	
		\subsubsection*{Asymptotic test for the HSIC}
		
Considering the HSIC, \citet{HSICtest} propose a kernel statistical test of independence based on asymptotic considerations.
	
\begin{theorem}[\citet{HSICtest}]\label{th:HSICtest}
Let the estimator $\text{HSIC}_n(X_k,Y)$ be rewritten
$$\text{HSIC}_n(X_k,Y)=\frac{1}{n^4}\sum_{i,j,q,r}^nh_{ijqr}$$
where $h_{ijqr}=\frac{1}{4!}\sum_{(t,u,v,w)}^{(i,j,q,r)}K_{\mathcal{X}_k,tu}K_{\mathcal{Y},tu}+K_{\mathcal{X}_k,tu}K_{\mathcal{Y},vw}-2K_{\mathcal{X}_k,tu}K_{\mathcal{Y},tv}$, the sum being done over the different permutations $(t,u,v,w)$ of $(i,j,q,r)$.

Then, under $\mathcal{H}_0$, the statistic $n\text{HSIC}_n(X_k,Y)$ converges in distribution to $\sum_{l>0}\lambda_lZ_l^2$, where the standard normal variables $Z_l$ are independent and where the coefficients $\lambda_l$ are the solutions of the eigenvalues problem $\lambda_l\psi_l(z_j)=\int h_{ijqr}\psi_l(z_i)dF_{iqr}$, $F_{iqr}$ being the distribution function of $(Z_i,Z_q,Z_r)$ and $\psi_l(.)$ the eigenvector associated to $\lambda_l$.
\end{theorem}

In practice \cite[for details, see][]{HSICtest}, the distribution of the infinite weighted sum of independent chi-squared variables is approached by a Gamma distribution with shape parameter $\gamma$ and inverse scale parameter $\beta$. The parameter $\gamma$ is estimated by
$\hat{\gamma}=\frac{n^{-2}(1+E_xE_y-E_x-E_y)^2}{V}$ and $\beta$ by $\hat{\beta}=\frac{nV}{n^{-1}(1+E_xE_y-E_x-E_y)}$ where:
\begin{itemize}
	\item $E_x=\frac{1}{n(n-1)}\sum_{1\leq i,j \leq n \atop i\neq j}\left(K_{\mathcal{X}_k}\right)_{ij}$, 
	\item $E_y=\frac{1}{n(n-1)}\sum_{1\leq i,j \leq n \atop i\neq j}\left(K_{\mathcal{Y}}\right)_{ij}$ and 
	\item $V=\frac{2(n-4)(n-5)}{n(n-1)(n-2)(n-3)}\mathbf{1}^T(B-\text{diag}(B))\mathbf{1}$, with $B=\left((HK_{\mathcal{X}_k}H)\odot(HK_{\mathcal{X}_k}H)\right)^{.2}$. $\odot$ is the element-wise multiplication and $M^{.2}$ the element-wise matrix po\-wer for all $M\in\mathcal{M}_n(\R)$.
\end{itemize}

Finally, the independence test rejects the null hypothesis $\mathcal{H}_0$ when the $p$-value of the Gamma distribution associated to the statistic $n\text{HSIC}_n(X_k,Y)$ is grea\-ter than some level $\alpha$, e.g. $\alpha=5\%$.

		\subsubsection*{Asymptotic test for the distance covariance}
	
For the distance covariance introduced in Section \ref{sec:dcov}, we refer to \citet{dcortest} and more precisely to the two following theorems:
\begin{theorem}[\citep{dcortest}]
If $\E[\|X_k\|_{d_k}+\|Y\|_p]<\infty$, then: 
\begin{itemize}
\item If $X_k$ and $Y$ are independent, $\frac{n\mathcal{V}_n^2}{S_2}\underset{n \rightarrow \infty}{\overset{\mathcal{L}}{\longrightarrow}} \sum_{l>0}\lambda_lZ_l^2$, where the standard normal variables $Z_l\sim\mathcal{N}(0,1)$ are independent and the $\lambda_l$ are positive reals.
\item If $X_k$ and $Y$ are dependent, $n\mathcal{V}_n^2/S_2\underset{n \rightarrow \infty}{\overset{\mathcal{P}}{\longrightarrow}} \infty$
\end{itemize}
where $S_2=\frac{1}{n^2}\left(\sum_{i,j=1}^nG^{(X_k)}_{ij}\right)\left(\sum_{i,j=1}^nG^{(Y)}_{ij}\right)$.
\end{theorem}

\begin{theorem}[\citet{dcortest}]\label{th:dcortest}
Let $T(X_k,Y,\alpha,n)$ be the statistical test rejecting the null hypothesis ``$\mathcal{H}_0$: $X_k$ are $Y$ independent'' when 
$$\frac{n\mathcal{V}_n^2}{S_2}>\left(\Phi^{-1}(1-\alpha/2)\right)^2$$
where $\Phi$ is the distribution function of the standard normal law and let $\alpha(X_k,Y,n)$ be its corresponding rate of type I error. 

If $\E[\|X_k\|_{d_k}+\|Y\|_p]<\infty$, then for all $\alpha \in ]0,0.215]$,
\begin{itemize}
\item $\lim_{n\rightarrow\infty}\alpha(X_k,Y,n)\leq \alpha$,
\item $\sup_{X_k,Y}\left\{\lim_{n\rightarrow \infty}\alpha(X_k,Y,n):\mathcal{V}(X_k,Y)=0\right\}=\alpha$.\\
\end{itemize}
\end{theorem}

Consequently, the test $T(X_k,Y,\alpha,n)$ has an asymptotic type I error rate at worst equal to $\alpha$ and the approximation of the $1-\alpha$ quantile of the law of $\sum_{l>0}\lambda_lZ_l^2$ by the squared $1-\alpha/2$ quantile of the standard normal law seems to be a powerful technique.
	
	\subsection{Spectral approach for the asymptotic tests}\label{sec:testspec}
	
For small and medium size samples, the previous approximations of the asymptotic laws are questionable. For example, \citet{dcortest} show that in the case of the distance covariance, the criterion presented in Theorem \ref{th:dcortest} might be over-conservative. In the context of a two-sample test, \citet{HSIC2samp} remind us of the heuristic nature of the Gamma approximation for the asymptotic law of the HSIC estimator. This substitution of laws can be not enough accurate for the upper tail of the distribution, that is to say for its most important part in the case of a $p$-value computation. Consequently, \citet{sejdinovic2013} advise the use of a spectrum approximation of the asymptotic laws for the HSIC and the distance covariance, which are weighted sums of chi-squares as mentioned in Theorems \ref{th:HSICtest} and \ref{th:dcortest}.\\

We approach the asymptotic law of $\frac{\text{HSIC}_n(X_k,Y)}{n}$ in Theorem \ref{th:HSICtest} by those of
$$\frac{1}{n^2}\sum_{i,j=1}^n\hat{\lambda}_{k,i}\hat{\nu}_j\varepsilon_{ij},\text{ with }\varepsilon_{ij}\overset{i.i.d.}{\sim}\mathcal{N}(0,1)$$
where $(\hat{\lambda}_{k,i})_{1\leq i \leq n}$ and $(\hat{\nu}_i)_{1\leq i \leq n}$ are the eigenvalues of $HK_{\mathcal{X}_k}H$ and $HK_{\mathcal{Y}}H$ respectively.\\

In the same way, we approach the asymptotic law of $n\mathcal{V}_n^2/S_2$ in Theorem \ref{th:dcortest} by the one of
$$\frac{1}{n^2}\sum_{i,j=1}^n\hat{\lambda}_{k,i}\hat{\nu}_j\varepsilon_{ij},\text{ with }\varepsilon_{ij}\overset{i.i.d.}{\sim}\mathcal{N}(0,1)$$
where $(\hat{\lambda}_{k,i})_{1\leq i \leq n}$ and $(\hat{\nu}_i)_{1\leq i \leq n}$ are the eigenvalues of $HG^{(X_k)}H$ and $HG^{(Y)}H$ respectively.\\

As it requires only the computation of the matrix-vector product $\lambda'\varepsilon_n\nu$ where $\lambda=(\lambda_1,\ldots,\lambda_n)'$, $\nu=(\nu_1,\ldots,\nu_n)'$ and $\varepsilon_n=(\varepsilon_{ij})_{1\leq i,j\leq n}$, an instance of such random variables appears clearly cheaper than a bootstrapped instance of the corresponding dependence measures. However this last approach can be required for small samples.
	
	\subsection{Non-asymptotic tests based on resampling}\label{sec:testboot}
	
The significance tests based on dependence measures presented in Sections \ref{sec:asymptest} and \ref{sec:testspec} are fast and asymptotically efficient tools for the selection of the influential input parameters. However, they are considerably biased when the number of observations $n$ is too weak because of their asymptotic framework. Consequently, non-asymptotic results are necessary.\\

For this purpose, we propose a generic non-pa\-ra\-me\-tric test based on resampling, which can be applied to any dependence measure $\Delta(X_k,Y)$ between two random variables $X_k$ and $Y$. For this, $B$ bootstrap versions $\mathbf{Y}^{[1]},\ldots,\mathbf{Y}^{[B]}$ of the output sample $\mathbf{Y}=(Y_1\ldots Y_n)$ are generated and for each $\mathbf{Y}^{[b]}$, the associated input sample is $\mathbf{X}^{[b]}_k:=\mathbf{X}_k$ where $\mathbf{X}_k=(X_{k,1}\ldots X_{k,n})$:
$$
 \underbrace{\begin{pmatrix}
  X_{k,1} & Y_1 \\
  \vdots  & \vdots  \\
  X_{k,n} & Y_n
 \end{pmatrix}}_{\text{Learning sample}}\longrightarrow
  \underbrace{\begin{pmatrix}
  X_{k,1} & Y_1^{[b]} \\
  \vdots  & \vdots  \\
  X_{k,n} & Y_n^{[b]}
 \end{pmatrix}_{1\leq b \leq B}.}_{\text{Bootstrap samples}}
$$

\begin{remark}Let us notice that the samplings are realized only with the output, all other things being equal. Indeed, our goal is to approach the distribution of the estimator $\Delta_n(X_k,Y)$ of $\Delta(X_k,Y)$ under the null hypothesis ``$X_k$ and $Y$ are independent'', thanks to many samplings of the sample $(\mathbf{X},\mathbf{Y})=\left(X_{1,i},\ldots,X_{d,i},Y_i\right)_{1\leq i \leq n}$. Consequently, as an infinity of $Y$ values can be associated to a particular value of $X_k$ under this hypothesis, any component of $\mathbf{Y}$ can be associated with probability $n^{-1}$ to the $i^{\text{th}}$ component of $\mathbf{X}_k$ in a bootstrap approach.
\end{remark}

Under these considerations, our test can be summarized by the following algorithm:

\begin{enumerate}
	\item Create a sample $(\mathbf{X},\mathbf{Y})=\left(X_{1,i},\ldots,X_{d,i},Y_i\right)_{1\leq i \leq n}$.
	\item Compute $\Delta_n(X_k,Y)$, an estimator of the dependence measure $\Delta(X_k,Y)$.
	\item Realize $B$ bootstrap samplings $(\mathbf{X}_k^{[b]},\mathbf{Y}^{[b]})$ of the sample $(\mathbf{X}_k,\mathbf{Y})$ under $\mathcal{H}_0$.
	\item Compute $\left(\Delta_n^{[b]}(X_k,Y)\right)_{1\leq b \leq B}$, the $B$ bootstrap estimators, where $\Delta_n^{[b]}(X_k,Y)$ is obtained by replacing $\mathbf{Y}$ by $\mathbf{Y}^{[b]}$ in the computation of $\Delta_n(X_k,Y)$.
	\item Compute the bootstrapped $p$-value $$p\text{-val}_B=\frac{1}{B}\sum_{b=1}^B\Ind_{\Delta_n^{[b]}(X_k,Y)>\Delta_n(X_k,Y)}.$$
	\item \underline{If} $p\text{-val}_B<\alpha$, \underline{then} reject $\mathcal{H}_0$, \underline{else} accept $\mathcal{H}_0$.\\
\end{enumerate}

\begin{remark}
This algorithm is designed for testing the dependence between an input parameter and the output of the model. If we want to simultaneously apply this test for the $d$ input parameters, only the steps 4 to 6 have to be repeated for each inputs. This avoids the repetition of the bootstrap step $d$ times.
\end{remark}

For methodological recommendations, we propose to use these significance tests based on resampling in the presence of a small sample and the use of the asymptotic tests presented in Section \ref{sec:asymptest} when the number of observations is much more important. Between both situations, spectral approach is better than the approximation of the asymptotic laws and it is also better than the use of the empirical distribution of a dependence measure estimator. Indeed, even if this last law is more justified than the asymptotic one, \citet{sejdinovic2013} highlight its important cost.  This is due to the computation of the dependence measure estimator for each bootstrapped sample, especially when the input or output parameter space dimension is important.

\section{Bootstrapped linear regression for the dependence measures}\label{sec:bootlinreg}

The previous significance tests for feature selection are directly computed on the dependence measures presented in Section \ref{sec:measures}, which associate one input parameter to the model output. In this section, we propose to decompose in a linear way the difference between two output observations according to the differences between the associated input observations; we call ``local measures'' these simple quantities measuring the difference between two observations of a same variable. Considering this linear model, our aim is to build significance tests for the different effects, using classical tools for nested model selection. Discarding an effect from this regression model corresponds to discarding a significant input parameter in a screening context.

	\subsection{Linear model between the local measures}\label{sec:linmod}

Considering a $n$-sample $\left(X_{1,i},\ldots,X_{d,i},Y_i\right)_{1\leq i \leq n}$ and a local measure $D(.,.)$, this linear model takes the form
\begin{equation}\label{eq:linmod}
D(Y_i,Y_j)=\sum_{k=1}^d\beta_kD(X_{k,i},X_{k,j}),~1\leq i,j\leq n
\end{equation}
where $\beta\in\R_+^d$. For two observations $(X_{1,i},\ldots,X_{d,i},Y_i)$ and $(X_{1,j},\ldots,X_{d,j},Y_j)$, the coefficient $\beta_k$ can be interpreted as the weight associated to the contribution of the dependence between $X_{k,i}$ and $X_{k,j}$ to the explanation of the dependence between $Y_i$ and $Y_j$.\\

The vector $\beta$ can be estimated by
\begin{equation}
\hat{\beta}\in\underset{\beta\in(\R_+)^{d}}{\arginf}\left\|D(\mathbf{Y})-\sum_{k=1}^d\beta_kD(\mathbf{X}_k)\right\|_{\text{Frob}}^2
\label{eq:minlinmod}
\end{equation}
where $\|.\|_{\text{Frob}}$ is the Frobenius norm defined for all $A\in\mathcal{M}_{n}(\R)$ by $\|A\|_{\text{Frob}}=\sqrt{\sum_{i,j=1}^nA_{ij}^2}$ and where $D(\mathbf{A})=\left(D(A_i,A_j)\right)_{1\leq i,j\leq n}$. As a function of the random variables $\mathbf{X}$ and $\mathbf{Y}$, $\hat{\beta}$ is also a random variable. For an easier implementation, we can rewrite Equation (\ref{eq:minlinmod}) with the Euclidean norm, replacing the matrix evaluations of the local measure $D$ by their vectorized forms:
\begin{eqnarray*}
\hat{\beta}\in&&\underset{\beta\in(\R_+)^{d}}{\arginf}\left\|\vec{D}(\mathbf{Y})-\sum_{k=1}^k\beta_k\vec{D}(\mathbf{X}_k)\right\|_2^2\\
&=&\underset{\beta\in(\R_+)^{d}}{\arginf}\left\|\vec{D}(\mathbf{Y})-\left[\vec{D}(\mathbf{X}_1)\ldots\vec{D}(\mathbf{X}_d)\right]\beta\right\|_2^2.
\end{eqnarray*}
where $\left(\vec{D}(\mathbf{Y})\right)_{(j-1)n+i}:=D(Y_i,Y_j),$  $\forall i,j\in\{1,\ldots,n\},$ and so on.

\begin{remark}In practice, the symmetric property of the matrices $D(\mathbf{Y}),~D(\mathbf{X}_1)$, ... and $D(\mathbf{X}_d)$ allows the use of smaller vectors $\vec{D}(\mathbf{Y}),~\vec{D}(\mathbf{X}_1)$, ... and $\vec{D}(\mathbf{X}_d)$ of size $\frac{n(n+1)}{2}$ instead of $n^2$.\end{remark}

\begin{remark}The decomposition of the $Y$ local measure into a linear combination of $X_1,\ldots,X_d$ local measures makes sense if the coefficients $\beta_1,\ldots,\beta_d$ are non-ne\-ga\-ti\-ve. This is the reason why the problem is a constrained linear least-squares minimization with $\beta\in(\R_+)^{d}$ rather than a simple linear least-squares minimization with $\beta\in\R^{d}$. This consideration leads to a more expensive problem resolution because of numerical optimization steps instead of an analytical solution $\hat{\beta}$.
 \end{remark}

The objective function in the constrained minimization problem (\ref{eq:minlinmod}) takes the form:
\begin{eqnarray*}
&&\eta(\mathbf{X},\mathbf{Y};\beta)=\left\|D(\mathbf{Y})-\sum_{k=1}^d\beta_kD(\mathbf{X}_k)\right\|_{\text{Frob}}^2
\\
&=&\Delta(\mathbf{Y},\mathbf{Y})-2\sum_{k=1}^d\beta_k\Delta(\mathbf{X}_k,\mathbf{Y})+\sum_{k,l=1}^d\beta_k\beta_l\Delta(\mathbf{X}_k,\mathbf{X}_l)
\end{eqnarray*}
where $\Delta(\mathbf{A},\mathbf{B})=\Tr\left[D(\mathbf{A})D(\mathbf{B})^T\right]\geq 0$. For certain local measures $D$, $\Delta$ quantifies the global dependence between the random variables $A$ and $B$ using $n$ independent evaluations stocked in $\mathbf{A}$ and $\mathbf{B}$. In these particular cases, this scheme is called ``minimal-redundancy-maximal-relevance'' (mRMR) because its minimization gives important weights to the input parameters maximizing the dependence measures $\Delta(\mathbf{X}_k,\mathbf{Y})$ and small weights to the input parameters highly dependent to the previous ones \citep{Peng05featureselection}. This can be very useful when many input parameters are dependent: in the extreme case where a parameter input is no more than a deterministic function of another one, we would be interested in a method keeping only one of these two variables.

Especially, in the case of the HSIC and using the notations of Section \ref{sec:HSIC}, the choices $D(\mathbf{Y})=HK_{\mathcal{Y}}H$ and $D(\mathbf{X}_k)=HK_{\mathcal{X}_k}H$, $\forall k\in\{1,\ldots,d\}$, lead to a result presented in \citet{daveiga}: 
\begin{eqnarray*}
&&\eta(\mathbf{X},\mathbf{Y};\beta)=\text{HSIC}_n(Y,Y)\\
&-&2\sum_{k=1}^d\beta_k\text{HSIC}_n(X_k,Y)+\sum_{k,l=1}^d\beta_k\beta_l\text{HSIC}_n(X_k,X_l).
\end{eqnarray*}

Likewise, in the case of the distance covariance and using the notations of Section (\ref{sec:dcov}), the choices $D(\mathbf{Y})=HG^{(Y)}H$ and $D(\mathbf{X}_k)=HG^{(X_k)}H$, $\forall k\in\{1,\ldots,d\}$, lead to the following mRMR scheme
\begin{eqnarray*}
&&\eta(\mathbf{X},\mathbf{Y};\beta)=\mathcal{V}_n^2(Y,Y)\\
&-&2\sum_{k=1}^d\beta_k\mathcal{V}_n^2(X_k,Y)+\sum_{k,l=1}^d\beta_k\beta_l\mathcal{V}_n^2(X_k,X_l).
\end{eqnarray*}

In a similar way, for the Pearson's coefficient correlation $\rho\left(X_k,Y\right)$, the choices of $D(\mathbf{Y})=AC$ and $D(\mathbf{X}_k)=B_kC$ with the matrices $A=\left(Y_i-Y_j\right)_{1\leq i,j \leq n}$, $B=\left(X_{k,i}-X_{k,j}\right)_{1\leq i,j \leq n}$ and $C=n^{-1}(n-1)^{-0.5}\mathbf{1}$ lead to the following mRMR scheme
\begin{eqnarray*}
&&\eta(\mathbf{X},\mathbf{Y};\beta)=\cov_n(Y,Y)\\
&-&2\sum_{k=1}^d\beta_k\cov_n(X_k,Y)+\sum_{k,l=1}^d\beta_k\beta_l\cov_n(X_k,X_l).
\end{eqnarray*} 

In the following, we consider an alternative to (\ref{eq:minlinmod}) for the estimation of the regression parameters of (\ref{eq:linmod}), particularly useful when the number of input parameters is important.
				
	\subsection{Shrinkage in high-dimension}\label{sec:lasso}
	
The coefficient estimation in the linear model (\ref{eq:linmod}) can be realized using regularization techniques. These methods are said active because they select the optimal complexity of the model during the optimization step (\ref{eq:minlinmod}) modified in some manner. More precisely, these techniques consist in the minimization of a quadratic risk penalized by an additive term, which is a constraint on the number or the size of model parameters, such as a limited $\ell^2$ norm \citep{ridge}, $\ell^1$ norm \citep{lasso} or a combination of both \citep{elasticnet}. In other words, a shrinkage tool looks for the optimal parameter values of (\ref{eq:linmod}) and the optimal effective dimension of the problem.

Under these considerations, we could use the Lasso (Least Absolute Shrinkage and Selection Operator) pe\-nal\-ty \citep{lasso} in order to select a subset of the local measures in the full model (\ref{eq:linmod}), and so a subset of the input parameters:

$$\eta_{\text{lasso}}(\mathbf{X},\mathbf{Y};\beta)=\left\|D(\mathbf{Y})-\sum_{k=1}^d\beta_kD(\mathbf{X}_k)\right\|_{\text{Frob}}^2+\lambda\|\beta\|_1.$$

It is in this sense that \citet{yamada} propose the HSIC Lasso which consists in the minimization of this objective function with $D(\mathbf{Y})=HK_{\mathcal{Y}}H$ and $D(\mathbf{X}_k)=HK_{\mathcal{X}_k}H$, $\forall k\in\{1,\ldots,d\}$, under the positivity constraints $\beta_1\geq 0,\ldots,\beta_d\geq 0$ and using a dual augmented Lagrangian algorithm to solve the optimization problem. 

In this paper, the HSIC Lasso is used but, for time computation reasons, we propose to solve the optimization problem with the Least Angle Regression (LARS) algorithm under positivity constraints \citep[Sec. 3.4]{LARS}, with a regularization parameter $\lambda$ optimized by an improved version of the cross-validation error minimization. Usually we take $\hat{\lambda}_{\text{CV}}$, the $\lambda$ value minimizing the cross-validation error $\mu_{\text{CV}}^{(l)}$. In the HSIC lasso framework, we propose to replace $\hat{\lambda}_{\text{CV}}$ by $\hat{\lambda}_{\text{CV mod}}$ which minimizes $$\mu_{\text{CV}}^{(l)}-0.5\sigma_{\text{CV}}^{(l)}$$
over the indices $\{1,\ldots,L\}$ of the discretized $\lambda$ values for the optimization, $\sigma_{\text{CV}}^{(l)}$ being the standard deviation of the prediction error associated to the different folds. The 0.5 value has been chosen after tests over various analytical functions. $\mu_{\text{CV}}^{(l)}-0.5\sigma_{\text{CV}}^{(l)}$ is an amelioration of the $\mu_{\text{CV}}^{(l)}$ objective to minimize because it takes into account the uncertainty of the cross-validation error. \\

Finally, if the effective dimension of the problem is of the same order than the number of input parameters, nested model selection tools can be considered instead of the shrinkage approach.

	\subsection{Bootstrap test for the nested model selection}\label{sec:bootlinmod}
	
Based on the full model (\ref{eq:linmod}) of Section (\ref{sec:linmod}), we propose some methods using significance tests in order to remove the non-significant input parameters. More precisely, for a given input parameter $X_k$, we want to build a statistical test with the null hypothesis ``$\mathcal{H}_0^{(k)}$: $X_k$ and $Y$ are independent'' and its alternative ``$\mathcal{H}_1^{(k)}$: $X_k$ and $Y$ are dependent'', or in a equivalent way: ``$\mathcal{H}_0^{(k)}$: $\beta_k=0$'' and ``$\mathcal{H}_1^{(k)}$: $\beta_k\neq 0$''. Obviously, the law of $\hat{\beta}_k$ in (\ref{eq:minlinmod}) is unknown and, at best, asymptotically approximable. Consequently, similarly to the resampling method proposed in Section \ref{sec:testboot}, we propose to build a bootstrap test for each input parameter $X_k$, starting from the $n$-sample $(X_{1,i},\ldots,X_{d,i},Y_i)_{1\leq i \leq n}$. More precisely, the $b^{\text{th}}$ bootstrap sample $(X^{[b]}_{1,i},\ldots,X^{[b]}_{d,i},Y^{[b]}_i)_{1\leq i\leq n}$ is such that 
$$Y^{[b]}_i:=Y_i,~X^{[b]}_{l,i}:=X_{l,i},~\forall l\neq k \text{ and } X^{[b]}_{k,i}:=X^{[b]}_{k,i}.$$
In other words, the $b^{\text{th}}$ bootstrap sample corresponds to the $n$-sample $(X_{1,i},\ldots,X_{d,i},Y_i)_{1\leq i \leq n}$  where the observations of the $k^{\text{th}}$ input parameter are resampled according to their empirical probability distribution. Then, we compute $\hat{\beta}^{[b]}$, the estimation of the vector $\beta$ for each bootstrap sample $(X^{[b]}_{1,i},\ldots,X^{[b]}_{d,i},Y^{[b]}_i)_{1\leq i\leq n}$.

Afterwards, under the null hypothesis $\mathcal{H}_0^{(k)}$, the $p$-value is estimated by
$$p\text{-val}_B^{(k)}=\frac{1}{B}\sum_{b=1}^B\Ind_{\hat{\beta}^{[b]}_k>\hat{\beta}_k}$$
and $\mathcal{H}_0^{(k)}$ is rejected when $p\text{-val}_B^{(k)}$ is lower than some level $\alpha$, e.g. $\alpha=5\%$.\\

Finally, considering the conclusions of the $d$ statistical tests, we obtain a sub-model of the full model (\ref{eq:minlinmod}) keeping only the significant local measures, and from another point of view, dismissing the non-significant input parameters of the model $Y=f(X_1,\ldots,X_d)$. From this conclusion, we could imagine to go further and to apply the tools commonly used for feature selection in the linear model, such as forward, backward or stepwise approaches.

\section{Numerical experiments}\label{sec:results}

In these sections, we numerically investigate the methods expounded in Sections \ref{sec:stattest} and \ref{sec:bootlinreg} for a screening purpose. We also compare the distance correlation and the HSIC with the classical Sobol' indices, in order to exhibit their specificities.

	\subsection*{Reminder on Sobol' indices}

For a model $Y(X)=f\left(X_1,\ldots,X_d\right)\in\R$ with independent random real variables $X_1,\ldots,X_d$ and such that $\E[f^2(X)]<+\infty$, we can apply the Hoeffding decomposition:
\begin{eqnarray*}
f(X)&=&f_0+\sum_{i=1}^df_j(X_j)+\sum_{i=1}^d\sum_{i < j}^df_{ij}(X_i,X_j)\\
&+&\ldots+f_{1\ldots d}(X_1,\ldots,X_d)=\sum_{u\subset\{1,\ldots,d\}}f_u(X_u)
\end{eqnarray*}
where $f_0=\E[f(X)]$, $f_j(X_j)=\E[f(X)|X_j]-f_0$ and $f_u(X_u)=\E[f(X)|X_u]-\sum_{v\subset u}f_v(X_v)$, with for all $u\subset\{1,\ldots,d\}$, $X_u=(X_i)_{i\in u}$. Then for each $u\subset\{1,\ldots,d\}$, the first-order and total Sobol' indices of $X_u$ are defined by
$$S_u=\frac{\int_{\mathcal{X}_u} f_u^2(x_u) d\mu_{X_u}(x_u)}{\int_{\mathcal{X}} f^2(x)d\mu_X(x)-f_0^2}\text{ and }S_u^T=\sum_{v\supset u}S_v,$$
where $\mu_{X_u}$ and $\mu_X$ are the distribution functions of $X_u$ and $X$ respectively. The first-order indices associated to $X_1,\ldots,X_d$ can also be rewritten: $S_k=\frac{\V\left[\E[f(X)|X_k]\right]}{\V[f(X)]}$, $\forall k\subset\{1,\ldots,d\}$.

	\subsection{Comparison of sensitivity indices}

These first tests on analytical functions aim at comparing various sensitivity indices: the classical Sobol' indices vs. dependence measures such as distance correlation (dCor), HSIC and sup-HSIC (the supremum of HSIC over the possible correlation length values. The objective is to identify which kinds of input effect they allow to detect, and to highlight any difference between these indices.

\begin{figure}[ht]
 \centering
  \includegraphics[width=0.4\linewidth]{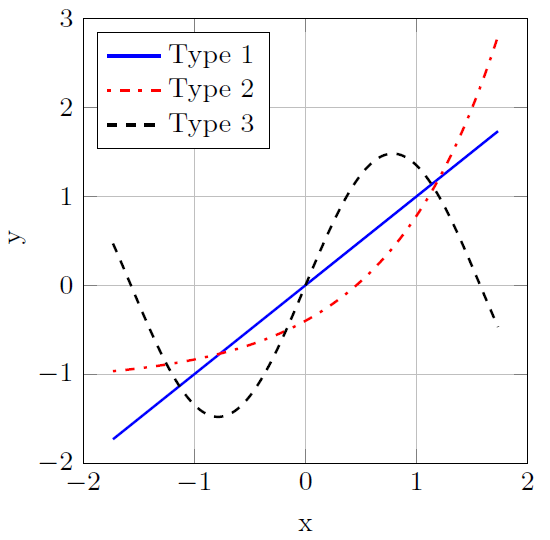}
	\caption{Elementary functions $h_{1}$, $h_{2}$ and $h_{3}$ on $[-\sqrt{3},\sqrt{3}]$.}	
	\label{fig:main_effects}
\end{figure}

For this, several analytical functions including linear or not, monotonic or not input effects are used in the following numerical tests. To build the different test functions, we considered monodimensional functions designed to be centered and with variance one when $x$ is a realization of an uniform random variable on $[-\sqrt{3},\sqrt{3}]$. These elementary functions are of three type:
\begin{enumerate}
\item linear: $h_{1}(x)=x$;
\item monotonous (exponential): $h_{2}(x)=\frac{e^{x}-a}{b}$ where $a=\frac{\sinh(\sqrt{3})}{\sqrt{3}}$ and $b=\sqrt{\frac{\sinh(2\sqrt{3})}{2\sqrt{3}}-a^2}$;
\item non-monotonous (sinusoidal): $h_{3}(x)=a\sin(2x)$ whe\-re $a=1/\sqrt{0.5-\frac{\sin(4\sqrt{3})}{8\sqrt{3}}}$.
\end{enumerate}
Each type of elementary function is represented in Figure \ref{fig:main_effects}.

The $d$ input parameters $X=\left(X_1,\ldots,X_d\right)$ of model $f$ are supposed independent and identically distributed according to an uniform distribution over $[-\sqrt{3},\sqrt{3}]$, such that these random variables are centered with variance one.

		\subsubsection*{Sensitivity indices regarding the shape of monodimensional effects}
		
First of all, we consider the additive model $f(X)$ with only monodimensional effects:
$$f(X)=\alpha_1h_{1}\left(X_1\right)+\alpha_2h_2\left(X_2\right)+\alpha_3h_3\left(X_3\right),~\alpha\in\R^3$$
and we propose to study the sensitivity of Sobol, HSIC, sup-HSIC and dCor indices to linear, monotonous and non-monotonous effects. Note that, in this case, the total Sobol' indices are equal to the first-order ones:
$$S_i=S_i^T=\frac{\alpha_i^2}{\alpha_1^2+\alpha_2^2+\alpha_3^2},~i\in\{1,2,3\}.$$ 
In the following tests, the coefficients $\alpha_i$ are set to 0 or 1, allowing to cancel the effect of the corresponding $X_i$.

The various sensitivity indices HSIC, sup-HSIC and dCor are estimated with a Monte-Carlo sampling of 1000 simulations, and compared to analytical Sobol index values. Note that, for this size of sampling, several Monte-Carlo repetitions have been performed and a negligible variance of dependence measure estimation has been observed, justifying  this choice of sample size. For each index, estimation is repeated 100 times. The mean values of sensitivity indices obtained for each kind of model are given in Table \ref{tab:meaning} in percentage (the sensitivity index for an input parameter is normalized by the sum of the sensitivity indices of different inputs). Firstly, the dependence measures HSIC, sup-HSIC and dCor are different from the Sobol' indices, with a relative difference up to 20\% with respect to the latter. Secondly, HSIC and sup-HSIC give the same results for these test functions. Then, the dependence measures give additional weight to linear effects, in comparison with the Sobol' indices. Finally, we observe differences between HSIC (and sup-HSIC) and dCor for non-linear functions, HSIC highligthing non-monotonous effects while dCor featuring monotonous ones.

\begin{table*}
	\centering
		\begin{tabular}{|c|c||c|c|c|c|}
			\hline
			$Y=f(X)=\ldots$ & Effect &  HSIC & sup-HSIC & dCor & \cellcolor{llightgray}Sobol\\
			\hline
			\hline
			\multirow{2}{*}{$h_1(X_1)+h_2(X_2)$} & $X_1$: linear & \textbf{62} & \textbf{61}  & \textbf{57} & \cellcolor{llightgray}50\\
			& $X_2$: monotonous & 38 & 39 & 43 & \cellcolor{llightgray}50 \\
			\hline 
			\multirow{2}{*}{$h_1(X_1)+h_3(X_3)$} & $X_1$: linear & \textbf{55} & \textbf{55}  & \textbf{63} & \cellcolor{llightgray}50 \\
			& $X_3$: non-monotonous & 45 & 45 & 37 & \cellcolor{llightgray}50 \\
			\hline
			\multirow{2}{*}{$h_2(X_2)+h_3(X_3)$} & $X_2$: non-linear & 44 & 45  & \textbf{56} & \cellcolor{llightgray}50 \\
			& $X_3$: non-monotonous & \textbf{56} & \textbf{55} & 44 & \cellcolor{llightgray}50 \\
			\hline
			\multirow{3}{*}{$h_1(X_1)+h_2(X_2)+h_3(X_3)$} & $X_1$: linear & \textbf{38} & \textbf{38} & \textbf{41} & \cellcolor{llightgray}33 \\
			& $X_2$: non-linear & 31 & 31 & 35 & \cellcolor{llightgray}33 \\
			& $X_3$: non-monotonous & 31 & 31 & 24 & \cellcolor{llightgray}33 \\
			\hline 
		\end{tabular}
		\caption{Sensitivity indices in percentage for different test functions.}
		\label{tab:meaning}
\end{table*}

		\subsubsection*{Sensitivity indices regarding the weight of the interaction effect}

Now, we consider the additive model $f(X)$ with a mo\-no\-di\-men\-sio\-nal and an interaction effect, the latter being weighted by a positive real $\alpha$:
$$f(X)=h_2\left(X_1\right)+\alpha h_2\left(X_1\right)h_2\left(X_2\right),~\alpha\in\R_+$$
and we propose to study the sensitivity of Sobol and HSIC to the value of $\alpha$. For brevity, we only present results for exponential shape, the conclusion being qualitatively the same for linear and sinusoidal ones. For the same reason, we only consider the dependence measure HSIC. The first-order Sobol' indices are $S_1=\frac{1}{1+\alpha^2}$ and $S_2=0$ while the total ones are equal to $S_1^T=1$ and $S_2^T=\frac{\alpha^2}{1+\alpha^2}$ for all $\alpha\in\R_+$.

\begin{table*}
	\resizebox{\linewidth}{!} {
	\begin{tabular}{|c||c|c||c|c||c|c||c|c||c|c|}
		\hline
		& \multicolumn{10}{c|}{$\alpha$} \\
		 & \multicolumn{2}{c|}{0} & \multicolumn{2}{c|}{1} & \multicolumn{2}{c|}{2} & \multicolumn{2}{c|}{4} & \multicolumn{2}{c|}{5} \\
		 \hline
		Measure & $X_1$ & $X_2$ & $X_1$ & $X_2$ & $X_1$ & $X_2$ & $X_1$ & $X_2$ & $X_1$ & $X_2$ \\
		\hline
		HSIC & \textbf{0.0965} & 0.0003 & 0.0293 & \textbf{0.0309} & 0.0071 & \textbf{0.0250} & 0.0092 & \textbf{0.0184} & 0.0104 & \textbf{0.0176} \\
		$\text{HSIC}_k/\sum_{j=1}^d\text{HSIC}_j$ & \textbf{99.7\%} & 0.3\% & 48.7\% & \textbf{51.3\%} & 22.2\% & \textbf{77.8\%} & 33.5\% & \textbf{66.5\%} & 37.2\% & \textbf{62.8\%} \\
		\hline
		Borgonovo $\delta_k$ & \textbf{0.7759} & 0.0044 & 0.4110 & \textbf{0.4530} & 0.2845 & \textbf{0.3993} & 0.2971 & \textbf{0.3610} & 0.3022 & \textbf{0.3546} \\
		$\delta_k/\sum_{j=1}^d\delta_j$ & \textbf{99.5\%} & 0.6\% & 47.8\% & \textbf{52.4\%} & 41.6\% & \textbf{58.4\%} & 45.1\% & \textbf{54.9\%} & 46.0\% & \textbf{54.0\%} \\		
		\hline
		\rowcolor{llightgray}Total Sobol $S^T_k$& \textbf{1} & 0 & \textbf{1} & 0.5000 & \textbf{1} & 0.8000 & \textbf{1} & 0.9412 & \textbf{}1 & 0.9615 \\
		\rowcolor{llightgray}$S^T_k/\sum_{j=1}^dS^T_j$ & \textbf{100\%} & 0\% & \textbf{66.7\%} & 33.3\% & \textbf{55.6\%} & 44\% & \textbf{51.5\%} & 48.5\% & \textbf{51.0\%} & 49.0\% \\
		\hline
	\end{tabular}
	}
	\resizebox{\linewidth}{!} {
	\begin{tabular}{|c||c|c||c|c||c|c||c|c||c|c|}
		\hline
		& \multicolumn{10}{c|}{$\alpha$} \\
		 & \multicolumn{2}{c|}{6} & \multicolumn{2}{c|}{7} & \multicolumn{2}{c|}{8} & \multicolumn{2}{c|}{9} & \multicolumn{2}{c|}{10}\\
		 \hline
		Measure & $X_1$ & $X_2$ & $X_1$ & $X_2$ & $X_1$ & $X_2$ & $X_1$ & $X_2$ & $X_1$ & $X_2$ \\
		\hline
		HSIC & 0.0112 & \textbf{0.0171} & 0.0118 & \textbf{0.0168} & 0.0123 & \textbf{0.0166} & 0.0127 & \textbf{0.0165} & 0.0130 & \textbf{0.0164}\\
		$\text{HSIC}_k/\sum_{j=1}^d\text{HSIC}_j$ & 39.6\% & \textbf{60.4\%}  & 41.3\% & \textbf{58.7\%} & 42.3\% & \textbf{57.4\%} & 43.5\% & \textbf{56.5\%} & 44.2\% & \textbf{55.8\%} \\
		\hline
		Borgonovo $\delta_k$ & 0.3059 & \textbf{0.3497} & 0.3086 & \textbf{0.3460} & 0.3106 & \textbf{0.3432} & 0.3121 & \textbf{0.3410} & 0.3133 & \textbf{0.3392}\\
		$\delta_k/\sum_{j=1}^d\delta_j$ & 46.7\% & \textbf{53.3\%} & 47.1\% & \textbf{52.9\%} & 47.5\% & \textbf{52.5\%} & 47.8\% & \textbf{52.2\%} & 48.0\% & \textbf{52.0\%}  \\
		\hline
		\rowcolor{llightgray}Total Sobol $S^T_k$ & \textbf{1} & 0.9730 & \textbf{1} & 0.9800 & \textbf{1} & 0.9846 & \textbf{1} & 0.9878 & \textbf{1} & 0.9901\\
		\rowcolor{llightgray}$S^T_k/\sum_{j=1}^dS^T_j$ & \textbf{50.7\%} & 49.3\% &  \textbf{50.5\%} & 49.5\% & \textbf{40.4\%} & 49.6\% & \textbf{50.3\%} & 49.7\% & \textbf{50.2\%} & 49.8\%\\
		\hline
	\end{tabular}}
	\caption{Standard and normalized HSIC, Borgonovo and total Sobol' indices for different values of $\alpha$.}
	\label{tab:inter}
\end{table*}

\begin{figure}[ht]
	\centering
	\includegraphics[width=0.6\linewidth]{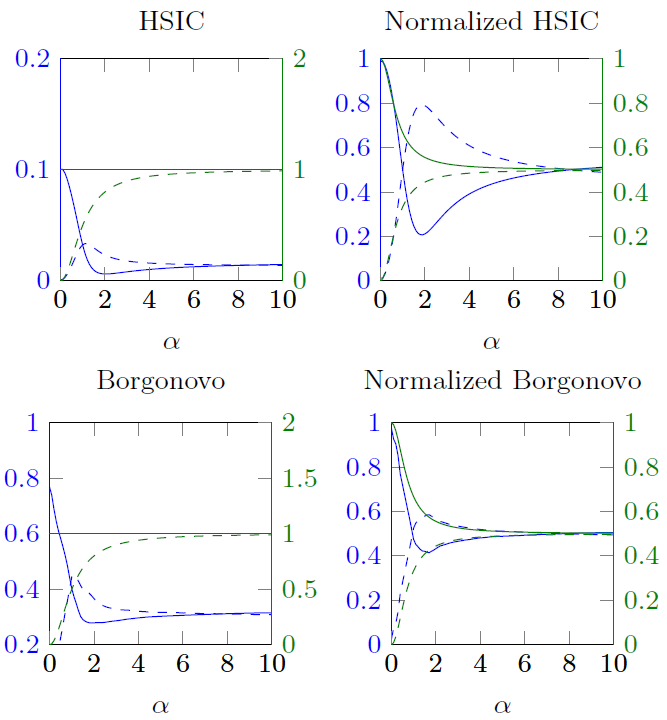}
	\caption{Standard and normalized HSIC and Borgonovo indices (blue lines) of $X_1$ (plain lines) and $X_2$ (dashed lines) for different values of $\alpha$, with the corresponding total Sobol' indices (green lines).}	
	\label{fig:inter}
\end{figure}

Table \ref{tab:inter} presents the mean HSIC estimations associated to this study for different values of $\alpha$ based on 1000 repetitions of a 1000-sample, while Figure \ref{fig:inter} illustrates the evolution of these indices according to the value of $\alpha$ for a certain 1000-sample. Naturally, the first variable is the more influential when $\alpha\ll 1$ because the second variable is almost missing in the model. Then, both variables tend to have the same effect around $\alpha=1$ and finally the second variable is the more influential with a pick around $\alpha= 2$ where the HSICs associated to $X_1$ and $X_2$ start a convergence to the same value. Table \ref{tab:inter} and Figure \ref{fig:inter} show that the same phenomenon occurs with the Borgonovo's delta moment independent measure \citep{Plischke2013536} defined by
$$\delta_k=\frac{1}{2}\int_{\mathcal{X}_k}f_{X_k}(x)\int_{\mathcal{Y}}\left|f_Y(y)-f_{Y|X_k=x}(y)\right|dy dx,$$
where $f_{X_k}$, $f_Y$ and $f_{Y|X_k=x}$ are the density probability functions of $X_k$, $Y$ and $Y|X_k$ respectively. Moreover, we have found the same result with the randomized dependence coefficient, a recent dependence measure defined in terms of correlation of random non-linear copula projections \citep{lopezpaz}. In the framework of the classical output variance decomposition, this situation is surprising: the second variable occurring only in the interaction effect, its total Sobol' index $S_2^T=\alpha^2(1+\alpha^2)^{-1}$ is obviously lower than the first input one $S_1^T=1$ for every $\alpha$. However, such situation is understandable if we look at the model $f$ from a multiplicative point of view, instead of an additional one:
$$f(X)=h(X_1)\left(1+\alpha h(X_2)\right).$$ 
This leads to the corresponding multiplicative decomposition of the variance:
$$\V\left[f(X)\right]=\E\left[\left(h\left(X_1\right)\right)^2\right]\times\E\left[\left(1+\alpha h\left(X_2\right)\right)^2\right]=1\times \alpha^2.$$
It appears that both inputs have the same contribution in the output variance when $\alpha=1$, $X_1$ is predominant when $\alpha<1$ and $X_2$ is predominant when $\alpha>1$. Moreover, when $\alpha$ tends to the infinity, the random function $f(X)$ tends to $\alpha h\left(X_2\right)h\left(X_1\right)$ where the effects of $X_1$ and $X_2$ on the output are completely equal because of the symmetry of the model $f$. 

\subsubsection*{Conclusion about dependence measures vs. Sobol' indices}

These particular analyses reveal that the HSIC and the distance correlation nuance the conclusion obtained with the Sobol' indices. While the latter only focus on the input parameter contribution in the output variance, the dependence measures seem to be more sensitive to the global behavior of the output. For the distance correlation, this can be explained by the fact that the distance covariance measures the distance between the product of the characteristic functions of a given input parameter and the model output and the characteristic function of the couple made by both variables. Consequently, it uses more information about the input-output relations because the characteristic function completely defines the probability distributions of these variables (separately and jointly). Furthermore, the HSIC maps the input and output values into the real line using some RKHS functions and measures the covariance between both functions; the associated estimator puts into relation the Gram matrices based on the associated reproducing kernels. In this way, the HSIC also uses more information about the output behavior than the Sobol' indices. Finally, GSA conclusions can be radically different between Sobol' and dependence measures in the presence of interaction effects.\\

In the next section, the dependence measures are considered in a screening framework for high-di\-men\-sio\-nal problems, where the number of influential input parameters is (much) lower than the total ones. In these situations, industrial applications often require to eliminate the non-significant inputs before the computation of the Sobol' indices for the significant ones, which are sensitivity indices of great interest for engineers. Indeed, the Sobol' approach robustness requires a lot of model evaluations, especially for high-dimension problems, and are not tractable for the whole set of input parameters; a subset of relevant inputs must be selected. Moreover, we have noted in numerical experiments not mentioned in this paper that, in the presence of influential and non-influential input parameters, the dependence measures take very different orders. Consequently, they arouse interest for screening purpose, in order to eliminate non-significant variables. It is in this sense we study the associated significance tests proposed and developed in Section \ref{sec:stattest}.

	\subsection{Significance tests for screening}
	
Now, we consider the function from \citet{morrisfunc} associating to the real input vector $$X=\left(X_1,\ldots,X_d,X_{d+1},\ldots,X_{d+\check{d}}\right)$$ the scalar output
\begin{equation}
f(X)=a\sum_{i=1}^d\left(X_i+b\sum_{i<j=2}^dX_iX_j\right)
\label{eq:morris}
\end{equation}
with $a=\sqrt{12}-6\sqrt{0.1(k-1)}$, $b=12\sqrt{0.1(k-1)}$ and $X_i\overset{i.i.d.}{\sim}\mathcal{U}\left([0,1]\right)$, $\forall i \in \{1,\ldots,d+\check{d}\}$. The $d$ first input parameters are the influential inputs while the $\check{d}$ are the non-influential ones. The ratio $r=\frac{\check{d}}{d}$ is the quantity of non-significant variables brought back to the quantity of significant ones.\\

The objective of this section is to evaluate the potential of the different dependence measures in terms of screening using their associated significant tests presented and proposed in Section \ref{sec:stattest}, for different sample sizes and different ratios $r$. A second objective is to study the screening performances of the Lasso regression and the bootstrap tests associated to the linear model (\ref{eq:linmod}) in Section \ref{sec:bootlinreg}.

		\subsubsection*{Comparison of different statistical tests based in dependence measures}

For this model $f$, asymptotic and bootstrap tests based on the dependence measures mentioned in Section \ref{sec:pearspea} are not at all satisfactory. Indeed, in the case of the Pearson's correlation coefficient with $n=500$ and $\check{d}=d$ for example, the null hypothesis is kept for all input parameters, with a mean $p$-value equal to 1 for each influential factors, and to 0.5 for the others. In the same way, we obtain a mean $p$-value equal to 0.5 for the Spearman's correlation  coefficient. This results are not surprising because the model $f$ is not linear and so does not respect the hypotheses underlying to these coefficients. This analytical application illustrates the limitations of such correlation coefficients and justifies the use of the other dependence measures such as HSIC and distance correlation. \\

Table \ref{tab:stattest} compares the significance tests associated to the HSIC and distance correlation using 1000 Monte-Carlo runs for each pair $(n,r)$ and computing the percentage of non-influential and influential input selection; the number of significant variables is equal to $d=5$. Among these last quantities, the first one is the rate of the type I error and the second one is the power of the test\footnote{The type I error occurs when the test concludes that a non-significant input is significant. The power of the test is the probability to conclude that a significant input is significant.}, usual notions in significance tests. Moreover, this table supplies the percentage of ``perfect screening'', which corresponds to the situation where all the non-significant variables are judged non-influential by the test while all the significant ones are judged influential.  Lastly, the significance tests are presented in their asymptotical (Section \ref{sec:asymptest}), spectral (Section \ref{sec:testspec}) and bootstrap (Section \ref{sec:testboot}) versions for each dependence measure with a level $\alpha$ equal to 5\%. \\

First of all, Table \ref{tab:stattest} shows that whatever the considered test, the rate of type I error and the power are independent of the non-significant input proportion $r$. Moreover, the rate of perfect screening increases with the number of observations $n$ and decreases with $r$. It is also higher with the distance correlation tests. We also note that a powerful test does not imply an important perfect screening rate. 

Then, considering the distance correlation, the results confirm the conservative property of the asymptotical test with a type I error around 1.5\%; this implies a test power lower than using the asymptotical test based on HSIC. However, the power increases with $n$ and this difference tends to disappear. Moreover, the bootstrap and the spectral tests give similar results, even if the spectral one is slightly conservative for a very small sample size while the bootstrap one is more powerful. Turning to HSIC, the asymptotical test has a type I error rate a little greater than the specified level (5\%) when $n$ is very small, while the spectral approach is slighly conservative. \\

Finally, we propose some advises to choose a statistical test according to the problem. A first point to note is the independence of the statistical tests to the proportion of non-significant variables; so, the remaining problem parameter is the number of observations. For a very small sample size, the bootstrap test for distance correlation is more powerful than the one based on HSIC. The distance correlation is also preferred to HSIC for a medium sample size for the same reason, but this time, the spectral approach is advised for the significance test. Finally, when the number of observations is sufficiently important (e.g. $n=200$ for $d+\check{d}=55$ variables), all the tests agree on conclusions, except the asymptotical test based on distance correlation which has a better rate of perfect screening because of its conservative aspect. So, asymptotical tests are the better solutions when $n$ is high, because of the previous reason for the distance correlation, and because of  economies in CPU time for both dependence measures.\\

Beyond the conclusions about the better approach (asymptotical, spectral or bootstrap), this comparison highlights that distance correlation is often more powerful than the HSIC. An explanation of this situation can be found in the definition of the distance covariance which measures the distance to the independence using characteristic functions, that is to say using the law definitions of the input parameters and of the output directly. However, the distance covariance is limited to vectorial inputs and outputs, contrarily to the HSIC which can deal with matricial inputs for example. Consequently, we advise to use the distance covariance for vectorial inputs and outputs and the HSIC for more complex data.
 	
 \begin{table}[ht]
  \centering
  \resizebox{\linewidth}{!}{
	\begin{tabular}{|c|c|c||c|c|c||c|c|c||c|c|c|}
		\hline
		& \multicolumn{2}{r||}{Significance test $\rightarrow$} & \multicolumn{3}{c||}{Asymptotical} & \multicolumn{3}{c||}{Spectral} & \multicolumn{3}{c|}{Bootstrap}  \\
		\cline{2-3}
		$n$ & & $r\rightarrow$ & 2 & 5 & 10 & 2 & 5 & 10 & 2 & 5 & 10   \\
		\hline
		\hline
		\multirow{6}*{10} & \multirow{2}*{Non-influential} & HSIC & 7.6 & 7.3 & 7.5 & 3.7 & 3.8 & 3.8 & 5.0 & 5.0 & 5.0\\
		 & & DCOR & \textbf{1.5} & \textbf{1.5} & \textbf{1.7} & 4.4 & 4.2 & 4.3 & 4.6 & 4.9 & 4.8 \\
		 \cline{2-12}
		 & \multirow{2}*{Influential} & HSIC & \textbf{19.8} & \textbf{20.1} & \textbf{21.1} & 13.1 & 14.1 & 13.6 & 15.3 & 16.3 & 16.2 \\
		 & & DCOR & 10.7 & 11.2 & 11.4 & \textbf{21.2} & \textbf{20.0} & \textbf{20.7} & \textbf{22.6} & \textbf{23.1} & \textbf{22.7} \\
		 \cline{2-12}
		 & \multirow{2}*{Perfect screening} & HSIC & 0.0 & 0.0 & 0.0 & 0.0 & 0.2 & 0.0 & 0.0 & 0.0 & 0.0\\
		 & & DCOR & 0.0 & 0.0 & 0.0 & 0.0 & 0.0 & 0.0 & 0.0 & 0.0 & 0.0 \\
		 \hline
		 \hline
		\multirow{6}*{25} & \multirow{2}*{Non-influential} & HSIC & 6.8 & 3.1 & 5.8 & 5.7 & 4.9 & 4.6 & 5.9 & 5.4 & 5.0 \\
			& & DCOR & \textbf{1.7} & \textbf{1.6} & \textbf{1.5} & 5.0 & 4.5 & 4.9 & 5.4 & 5.1 & 4.7 \\
			\cline{2-12}
		 & \multirow{2}*{Influential} & HSIC & \textbf{40.1} & \textbf{40.1} & \textbf{40.4} & 37.7 & 37.0 & 37.5 & 38.5 & 38.6 & 38.7 \\
		 & & DCOR & 37.9 & 37.4 & 37.7 & \textbf{56.1} & \textbf{56.3} & \textbf{56.9} & \textbf{57.7} & \textbf{57.3} & \textbf{57.9} \\
		 \cline{2-12}
		 & \multirow{2}*{Perfect screening} & HSIC & 0.3 & 0.4 & 0.0 & 0.2 & 0.2 & 0.0 & 0.3 & 0.3 & 0.0 \\
			& & DCOR & 0.2 & 0.2 & 0.1 & 3.0 & 2.0 & 0.0 & \textbf{2.6} & \textbf{2.6} & 0.1 \\		 
			\hline
		 \hline
		\multirow{6}*{50} & \multirow{2}*{Non-influential} & HSIC & 5.3 & 5.2 & 5.3 & 4.6 & 4.6 & 4.7 & 5.0 & 4.8 & 4.9 \\
			& & DCOR & \textbf{1.1} & \textbf{1.3} & \textbf{1.4} & 4.7 & 4.6 & 5.0 & 4.6 & 4.5 & 4.7 \\
			\cline{2-12}
		 & \multirow{2}*{Influential} & HSIC & 70.8 & 70.0 & 71.5 & 69.2 & 68.4 & 70.1 & 69.9 & 68.8 & 70.4 \\
		 	& & DCOR & \textbf{75.4} & \textbf{74.6} & \textbf{76.7} & \textbf{87.5} & \textbf{87.8} & \textbf{87.7} & \textbf{88.4} & \textbf{87.7} & \textbf{89.2} \\
		 	\cline{2-12}
		 & \multirow{2}*{Perfect screening} & HSIC & 9.7 & 0.8 & 0.5 & 8.9 & 8.0 & 0.7 & 9.9 & 7.9 & 0.5 \\
		 	& & DCOR & \textbf{19.3} & \textbf{17.2} & \textbf{9.4} & \textbf{35.6} & \textbf{29.6} & \textbf{4.0} & \textbf{41.4} & \textbf{30.1} & \textbf{4.7}\\
		 	\hline
		 \hline
		\multirow{6}*{100} & \multirow{2}*{Non-influential} & HSIC & 5.1 & 5.3 & 5.4 & 4.9 & 4.9 & 5.1 & 4.9 & 4.8 & 5.0 \\
			& & DCOR & \textbf{1.6} & \textbf{1.3} & \textbf{1.6} & 4.4 & 4.7 & 5.1 & 4.9 & 4.7 & 5.0 \\
			\cline{2-12}
		 & \multirow{2}*{Influential} & HSIC & 95.7 & 95.9 & 95.9 & 95.5 & 95.8 & 95.7 & 95.6 & 95.7 & 95.8 \\
		 	& & DCOR & \textbf{98.0} & \textbf{98.2} & \textbf{98.0} & 99.5 & 99.6 & 99.5 & \textbf{99.4} & \textbf{99.5} & \textbf{99.4}\\
		 	\cline{2-12}
		 & \multirow{2}*{Perfect screening} & HSIC & 61.1 & 47.3 & 6.2 & 60.7 & 49.0 & 6.8 & 61.4 & 49.0 & 6.9 \\
		 	& & DCOR & \textbf{83.9} & \textbf{80.8} & \textbf{40.7} & \textbf{77.9} & \textbf{60.2} & \textbf{9.1} & \textbf{75.5} & \textbf{60.1} & 6.9\\
		 \hline
		 \hline
		\multirow{6}*{200} & \multirow{2}*{Non-influential} & HSIC & 4.7 & 5.3 & 5.2 & 4.5 & 5.0 & 4.9 & 4.5 & 4.9 & 5.0 \\
			& & DCOR & \textbf{1.3} & \textbf{1.2} & \textbf{1.4} & 4.2 & 5.2 & 5.1 & 4.7 & 5.0 & 5.0\\
			\cline{2-12}
		 & \multirow{2}*{Influential} & HSIC & 99.9 & 99.9 & 100 & 99.9 & 99.9 & 100.0 & 99.9 & 99.9 & 100 \\
		 	& & DCOR &  100 & 100 & 100 & 100 & 100 & 100 & 100 & 100 & 100\\
		 	\cline{2-12}
		 & \multirow{2}*{Perfect screening} & HSIC & 78.5 & 57.5 & 6.1 & 79.3 & 59.6 & 7.5 & 79.1 & 59.8 & 7.4 \\
		 	& & DCOR & \textbf{93.4} & \textbf{88.5} & \textbf{47.5} & 80.8 & 59.2 & 8.3 & 78.8 & 60.1 & 7.2\\
		 \hline
	\end{tabular}}
	\caption{Percentage of non-influential and influential input selection and perfect screening for different 5\%-level significance tests, different sample sizes and different ratios of non-influential inputs, with HSIC and dCor. }
	\label{tab:stattest}
\end{table}

		\subsubsection*{Linear regression with HSIC}

In this second part, we apply the linear model (\ref{eq:linmod}) based on HSIC to the numerical model (\ref{eq:morris}) for $d=\check{d}=5$. We consider the bootstrap significance tests proposed in Section \ref{sec:bootlinmod} for the nested model selection and the HSIC Lasso with cross-va\-li\-da\-tion proposed in Section \ref{sec:lasso}. Table \ref{tab:linmod} gives the percentage of selected non-influential and influential inputs as well as the percentage of perfect screening, for different sample sizes and different methods: the bootstrap significance tests with 5\%-level (\texttt{bootstrap}), the HSIC Lasso with cross-validation minimization (\texttt{Las\-so 1}) and the HSIC Lasso with our improved cross-validation minimization (\texttt{Las\-so 2}). $N=1000$ Monte-Carlo runs have been realized and, for the Lasso regression, we have adapted a Matlab implementation of the LARS algorithm\footnote{Matlab implementation of the LARS algorithm: http://www.stat.berkeley.edu/$\sim$yugroup/downloads/.} for the positive Lasso \citep{LARS}. Firstly, whatever the number of observations $n$, the bootstrap approach selects no more than 5\% of non-influential inputs while the Lasso methods keep a lot of these variables: more than 60\% with \texttt{Lasso 1} and between 8 and 20\% with \texttt{Lasso 2}, according to the sample size. On the contrary, the bootstrap approach is less powerfull with small samples than the Lasso regression. Finally, the HSIC Lasso with our improved version of cross-validation minimization leads to a better perfect screening rate than using the classical one and when the ratio $n/(d+\check{d})$ increases, the boostrap approach is the method providing the more accurate screening. 

To conclude, the bootstrap significance tests proposed in Section \ref{sec:bootlinmod} for the nested model selection constitute the best approach for screening in a linear regression framework, except if $n/(d+\check{d})$ is too small. Moreover, HSIC Lasso is an interesting tool but the choice of the penalization constant $\lambda$ is an open-pro\-blem. Numerical tests reveal that its selection by cross-va\-li\-da\-tion minimization keeps a too important number of non-influential variables, while our improved version seems to be a promising alternative. Lastly, we remark that the bootstrap results are less good than those using the bootstrap distribution of the HSIC under the null hypothesis in Table \ref{tab:stattest}.

\begin{table}[ht]
	\centering
	\begin{tabular}{|c|c|c|c|c|}
	\hline
		$n$ & & \texttt{Bootstrap} & \texttt{Lasso 1} & \texttt{Lasso 2} \\
		\hline
		\multirow{3}*{50}	& Non-influential & \textbf{4.0} & 63.9 & 20.1 \\
		& Influential & 68.2 & \textbf{97.7} & 85.7 \\
		& Perfect screening & 7.4 & 1.4 & \textbf{14.9} \\
		\hline
		\multirow{3}*{100}	& Non-influential & \textbf{4.2} & 66.3 & 14.1 \\
		& Influential & 92.5 & \textbf{99.9} & 96.1 \\
		& Perfect screening & \textbf{48.6} & 0.7 & 40.9 \\
		\hline
		\multirow{3}*{200}	& Non-influential & \textbf{4.5} & 62.3 & 8.2 \\
		& Influential & 99.8 & 100 & 99.6 \\
		& Perfect screening & \textbf{74.9} & 2.6 & 66.6 \\		
		\hline
	\end{tabular}
	\caption{Percentage of non-influential and influential input selection and perfect screening for different feature selection approaches and different sample sizes, with HSIC in the linear regression model. }
	\label{tab:linmod}
\end{table}	

\section{Conclusion}

In this paper, we introduce new developments around the use of dependence measures for sensitivity analysis (SA) and screening purposes. This situation occurs notably during the first steps of the establishment of a model, when the influential inputs are not exactly known and the precaution requires to consider all the potentially significant variables. Because of the costly nature of the numerical simulator, only some observations can be obtained, which prevents the use of classical SA quantitative methods, such as Sobol' indices, for these high-dimensional problems. Furthermore, classical Sobol' indices only focus on the decomposition of the output variance and not on its entire probabilistic distribution. For all these reasons, we turn to dependence measures recently proposed for global sensitivity analysis: the distance correlation and the Hilbert-Schmidt independence criterion (HSIC). The HSIC considers the covariance between two RKHS functions applied to theses variables, and the distance covariance leading to the distance correlation corresponds to the mean norm between the characteristic function of both variables and the product of the characteristic functions of these variables. \\

At first, considering a sparse problem where the number of non-significant input parameters can be very important, independence hypothesis tests are required to use these new measures directly for a screening purpose. For this, asymptotic versions of such tests exist. Spectral approximations for the probabilistic laws involved in the asymptotic tests could improve some intrinsic approximations, especially in the presence of a medium size sample. From this, we propose non-asymp\-to\-tic versions for these independence tests, in the case where the number of observations is low compared to the number of uncertain inputs. These non-asymptotic tests are based on a bootstrap sampling method. Always for a screening purpose, we propose a second approach based on the decomposition of any local measure of difference between two observed outputs as a linear regression on the same measures between the corresponding inputs. The regression coefficients are estimated using a linear least-squares minimization under positivity constraints. A coefficient equal to zero means that the corresponding input has no significant influence on the output. Thus, testing the nullity of each coefficient provides  a screening method. In the case of the HSIC, we show that this model with $\ell^1$-penalization corresponds to the HSIC Lasso approach for feature selection and we propose to solve this problem using the LARS algorithm with positive coefficients. We also introduce a method for the selection of the penalty constant, based on the minimization of the cross-validation error reduced by a weighting of the associated standard-deviation. Likewise, we propose to apply the classic tools of model selection and, in particular, a bootstrap method testing the nullity of the model coefficients. To compare the different proposed approaches for screening based on dependence measures, we performed several numerical tests on classical analytical functions. Concerning the first approach, these experiments show that the different proposed significant tests based on dependence measures are very efficient. The ones based on distance correlation are sometimes more powerful while the ones based on HSIC have the advantage to be well-adapted to the case of high dimensional inputs. Concerning the kind of significance test (asymptotic, spectral and non-asymptotic), the compromise ``CPU time - accuracy'' gives the advantage to the bootstrap tests in the presence of small sample sizes and to the asymptotical approaches when the number of observations is higher. The spectral approximation of the asymptotical law can be viewed as an intermediary solution between these two extreme configurations. In addition,  the first approach using directly the dependence measures seems to be slightly better than the second one, based on the linear decomposition of these sensitivity measures.\\

In this paper, we also try to provide some preliminary answers to the question ``What sensitivity indices to what situation?'', without pretension to build a theory. For this, we performed many tests on toy functions to compare the results given by the HSIC, the distance correlation and the Sobol index. Firstly, the new dependence measures lead to conclusions of the same order than the Sobol' indices ones. Then, they seem to be higher for the linear effects than for the non-linear ones, these effects being additive, centered and with variance equal to one, which leads to uniform Sobol' indices. Moreover, the HSIC further detects the monotonic effects v.s. the non-monotonic ones while the opposite occurs with the distance correlation. These tests also highlights that the dependence measures are more sensitive to the presence of an interaction term than the Sobol' indices and yield some different sensibility analysis conclusions. Their interpretation seems closer to that of the density-based sensitivity indices such as Borgonovo's. Beyond this complementary aspect, the HSIC and the distance correlation need only a few number of model evaluations, which is a great advantage over the classical variance-based or density-based indices. Finally, various numerical tests illustrate that dependence measures provide a relevant information which is coherent and sometimes complementary to the one obtained with classical indices.\\

Given the above, we advise the use of dependence measures associated to independence tests in global sensitivity analysis when the number of simulations is weak, when the problem takes place in an high-dimensional context or when we want to reinforce or qualify the conclusions obtained with the Sobol' indices. Moreover for industrial problems, the aim of the GSA is often to reduce the output variance of the simulator. In this case, the use of dependence measures can be viewed as a selection step and then, a quantitative phase consists to  compute the Sobol' indices of the retained model inputs. \citet{daveiga} also shows the interest of such sensitivity measures for high-dimensional output. However, in the presence of many thousands of observations, distance correlation and HSIC estimations are CPU time-expensive and other kernel methods should be investigated. More recently, a new dependence measure called ``randomized dependence coefficient'' has been proposed \citep{lopezpaz} with a computational cost of $\mathcal{O}\left(n\log(n)\right)$ while the distance correlation and HSIC ones are of $\mathcal{O}\left(n^2\right)$, $n$ being the number of required simulations. Considering this coefficient for GSA problems could be an interesting extension to this paper. In addition, applying the present screening methods to industrial applications, with functional inputs and outputs, could be a follow-up to this work. Finally, it should be interesting to study the dependence of the significance test results to the kernel functions chosen to compute the HSIC.

\section*{Acknowledgments}

We are grateful to B\'eatrice Laurent and S\'ebastien Da Veiga for helpful discussions.
	
\bibliographystyle{spbasic}      

\end{document}